\documentclass[dvips]{acta}
\usepackage{supertabular,lscape,epsfig}
\usepackage{amssymb}
\usepackage{amsmath}
\SetPages{1}{7}
\SetVol{60}{2010}

\begin{document}

\begin{Titlepage}

\Title { On the Mass Transfer Rate in SS Cyg }

\Author {J.~~S m a k}
{N. Copernicus Astronomical Center, Polish Academy of Sciences,\\
Bartycka 18, 00-716 Warsaw, Poland\\
e-mail: jis@camk.edu.pl }

\Received{  }

\end{Titlepage}

\Abstract { The mass transfer rate in SS Cyg at quiescence, estimated 
from the observed luminosity of the hot spot, is $\log \dot M_{tr}=16.8\pm0.3$.  
This is safely below the critical mass transfer rates of 
$\log \dot M_{crit}=18.1$ (corresponding to $\log T_{crit}^\circ=3.88$) 
or $\log \dot M_{crit}=17.2$ (corresponding to the "revised" value 
of $\log T_{crit}^\circ=3.65$). 
The mass transfer rate during outbursts is strongly enhanced.  
} 
{accretion, accretion disks -- binaries: cataclysmic variables, stars: 
dwarf novae, stars: individual: SS Cyg }

\section { Introduction } 

SS Cyg is one of the best studied dwarf novae. Its system parameters 
have recently been reliably determined by Bitner et al. (2007). 
Earlier Harrison et al. (1999) determined the new, highly accurate trigonometric 
parallax $\pi=(6.02\pm 0.46)$ mas, corresponding to a distance of 
$d=166\pm 12$ pc, i.e. 1.5-2.0 times larger than estimated previously. 
 
Schreiber and G{\"a}nsicke (2002) discussed the consequences of this 
new distance and found that the mass transfer rate in SS Cyg at quiescence 
is significantly {\it larger} than the critical mass transfer rate. 
Their analysis was later repeated with new system parameters (from 
Bitner et al. 2007) by Schreiber and Lasota (2007) who found 
$\dot M_{tr}=(8.8-9.2)\times 10^{18}$ g/s, compared to 
$\dot M_{crit}=(9.2-9.1)\times 10^{17}$ g/s. This discrepency led them 
to the conclusion that {\it "Either our current picture of disc accretion 
in these systems must be revised or the distance to SS Cyg is $\sim$ 100 pc"}. 

Common to both these investigations was the assumption that the mass transfer 
rate at quiescence can be calculated as the amount of mass accreted 
during outburst (estimated from the absolute visual magnitude at outburst 
maximum and its duration) divided by the length of the cycle. 
This assumption would be correct only in the case of a constant mass transfer 
rate throughout the entire dwarf nova cycle. 
However, as already pointed out by Schreiber and Lasota, there is growing 
evidence that during outbursts and superoutbursts of dwarf novae the mass 
transfer rate is strongly enhanced. 

The mass transfer rate during quiescence can be determined directly from 
the luminosity of the hot spot. This will be done in the present paper.

\section { System Parameters and the Critical Mass Transfer Rate } 

\subsection {System Parameters}

We adopt system parameters of SS Cyg determined by Bitner et al. (2007). 
These are: $M_1=0.81\pm0.19~M\odot$, $M_2=0.55\pm0.13~M\odot$, and 
$i=49^{+6}_{-4}$ (note that $i=49$ corresponds to $M_1=0.81$ 
and $M_2=0.55$). 

The mean magnitude and color of SS Cyg at quiescence compiled by 
Bruch and Engel (1994) are: $V=11.9$, $(B-V)=0.53$. After being corrected 
for $E_{B-V}=0.07$ they become: $V_\circ=11.7$, $(B-V)_\circ=0.46$. 
Using distance modulus of $(m-M)=6.10\pm0.15$, corresponding to the 
trigonometric parallax measured by Harrison et al. (1999), we get 
$M_V=5.6\pm0.15$. In addition, however, we must note that -- due to variations 
of the V magnitude at quiescence -- the uncertainty of this value is likely 
to be bigger. In what follows we will adopt: $M_V=5.6\pm0.5$. 

\subsection {The Hot Spot in SS Cyg at Quiescence}

Voloshina and Khruzina (2000a; English translation: 2000b) 
published mean UBV light curves of SS Cyg 
at quiescence showing double-humped modulation due -- obviously -- 
to the non-spherical shape of the secondary. The two maxima are of nonequal 
heights, the one at phase $\phi\sim 0.75$ being higher. 
This was interpreted by them and also by Bitner et al. (2007), as being due 
to the contribution from the hot spot. 

Adopting this intepretation we determine the absolute visual magnitude of 
the spot at its maximum. Using the observed difference between the amplitudes 
of two maxima $\Delta f_V/<f_V>\approx 0.047\pm 0.005$ (see Fig.2 in Voloshina 
and Khruzina 2000ab, or Fig.4 in Bitner et al. 2007) and $M_V=5.6\pm0.5$ 
(as adopted above) we obtain $M_{V,sp}^{max}=8.9\pm0.5$. 

The B and V light curves (Fig.2 in Voloshina and Khruzina 2000ab) show nearly 
identical amplitudes. This implies that the color of the hot spot 
is: $(B-V)_{sp}\approx (B-V)_\circ=0.46$. 
Using calibration based on Kurucz (1993) model atmospheres we get for the 
effective temperature of the spot: $\log T_{sp}\approx 3.82$.

\subsection {The Critical Mass Transfer Rate}

The critical mass transfer rate, below which the thermal instability sets 
in and results in the dwarf nova behavior, is defined by the 
critical temperature corresponding to the upper bend in the $\Sigma-T_e$ 
relation (see Lasota 2001 or Smak 2002 and references therein). 
The value of this critical temperature depends on relevant parameters as  
(Smak 2002, Eq.13) 

\beq
\log T_{crit}~=~\log T_{crit}^\circ~-~0.085~\log R_{d,10}~+~
      {{0.085}\over {3}} \log M_1~+~0.010~ \log (\alpha/0.1)~,
\eeq

\noindent
where $M_1$ is in solar units, $R_{d,10}=R_d/10^{10}$ and $T_{crit}^\circ$ 
corresponds to $M_1=1$, $R_d=1\times 10^{10}$ and $\alpha=0.1$. 

The critical mass transfer rate is then related to the critical 
temperature by 

\beq
\sigma~ T_{crit}^4~=~{{3}\over {8\pi}}~\dot M_{crit}~{{GM_1}\over {R_d^{~3}}}~
        \left[ 1~-~(R_1/R_d)^{1/2} \right]~.
\eeq 

The commonly adopted value of the critical temperature 
$\log T_{crit}^\circ \approx3.88$ is based on the shape of the $\Sigma-T_e$ 
relations resulting from numerical integrations of the vertical structure 
equations. However, there is evidence (Smak 2002), based on the analysis 
of CV's with stationary accretion and, particularly, of Z Cam stars at 
their standstills, which strongly suggest that its value is much lower, 
namely: $\log T_{crit}^\circ\approx3.65$. 

Using system parameters of SS Cyg and $r_d=R_d/A=r_{tid}=0.9~r_{Roche}$  
we get for those two values of $T_{crit}^\circ$

\beq
\log~\dot M_{crit}^{3.88}~=~18.1~,~~~~~{\rm and}~~~~~
\log~\dot M_{crit}^{3.65}~=~17.2~.
\eeq

\section { Mass Transfer Rate at Quiescence from the Hot Spot }

\subsection { Luminosity of the Hot Spot }

The mean bolometric luminosity of the hot spot can be written as 

\beq
<L_{bol,sp}>~=~{1\over 2}~\Delta V^2~\eta~\dot M_{tr}~=
       ~{1\over 2}~\Delta v^2~\left({{2\pi}\over{P}}A\right)^2\dot M_{tr}~,
\eeq 

\noindent
where $1/2\Delta V ^2$ is the energy dissipation per 1 gram of the stream
material, $\dot M_{tr}$ is the mass transfer rate, and $\Delta v^2$ is the 
dimensionless equivalent of $\Delta V^2$ which depends on the mass ratio 
and on the distance from the disk center, i.e. on the radius of the disk 
$r_d=R_d/A$ or $r_d/r_{Roche}$. 

In general, the luminosity of the hot spot represents only part of the 
energy dissipated during stream collision. 
To account for this effect Eq.(4) contains factor $\eta<1$. 
In what follows we will adopt $\eta=0.5$. This is probably a {\it lower limit} 
to the true value of this parameter. If so, the resulting value of 
$\dot M_{tr}$ will form an {\it upper limit} to the true mass transfer rate, appropriate in the context of our considerations. 

The luminosity of the spot can also be expressed as

\beq
<L_{bol,s}>~=~\pi~s^2~A^2~\sigma~T_{sp}^4~,
\eeq

\noindent
where $s$ is the dimensionless radius of the spot (assumed to be circular) 
and $T_{sp}$ -- its effective temperature. 
By comparing Eqs.(4) and (5) we obtain $s$ as a function of other paremeters 

\beq
s~=~\left [ { 
{ {1\over 2}~\Delta v^2~\eta~(2\pi/P)^2~\dot M_{tr} }
\over {\pi~\sigma~T_{sp}^4} } \right ]^{1/2}~.
\eeq

The mean {\it visual} luminosity of the spot can now be calculated as  

\beq
<L_{V,sp}>~=~\pi~s^2~A^2~f_V(T_{sp})~,
\eeq

\noindent
where $f_V(T_{sp})$ is the visual flux, to be obtained from Kurucz (1993) 
model atmospheres. 

Turning to the maximum luminosity of the spot observed at orbital 
inclination $i$ we have (Paczy{\'n}ski and Schwarzenberg-Czerny 1980, 
see also Smak 2002)

\beq
L_{V,sp}^{max}~=~{{12}\over {3-u}}~(1~-u~+~u\sin i)~\sin i~<L_{V,sp}>~,      
\eeq

\noindent
where $u$ is the limb darkening coefficient. 
Converting $L_{V,sp}^{max}$ to magnitudes we finally obtain  

\beq
M_{V,sp}^{max}~=~f(\dot M_{tr},r_d/r_{Roche},T_{sp},u)~. 
\eeq

Calculations show that -- with parameters applicable to SS Cyg -- the 
largest uncertainty in the resulting $M_{V,sp}^{max}-\dot M_{tr}$ (Eq.9) 
and $s-\dot M_{tr}$ (Eq.6) relations comes from $r_d$. 
Using results obtained from spot eclipse analysis in U Gem (Smak 2001) 
and IP Peg (Smak 1996) we adopt: $r_d=(0.7\pm0.1)~r_{Roche}$. 
For the limb darkening coefficient we adopt $u=0.6$; additional calculations 
with $u=0.2$ show that $M_{V,sp}^{max}$ is rather insensitive to this parameter. 
The effects of $T_{sp}$ will be discussed below. 

The $M_{V,sp}^{max}-\dot M_{tr}$ relation depends of course 
also on system parameters. Fortunately, it turns out that in this case 
the effects of higher masses at lower inclination (and vice versa) largely 
compensate each other. 

The resulting $M_{V,sp}^{max}-\dot M_{tr}$ and $s-\dot M_{tr}$ relations 
are shown in Fig.1. 
In addition to relations calculated with $\log T_{sp}=3.82$ (Section 2.2) 
shown are also relations corresponding to $\log T_{sp}=4.00$; 
note that such a temperature would already require the spot to have 
$(B-V)\approx 0.0$, i.e. be bluer than observed by nearly 0.5 mag. 
Even in such a case the $M_{V,sp}^{max}-\dot M_{tr}$ relation differs 
only slightly from the relation calculated with $\log T_{sp}=3.82$. 

\begin{figure}[htb]
\epsfysize=9.0cm 
\hspace{0.1cm}
\epsfbox{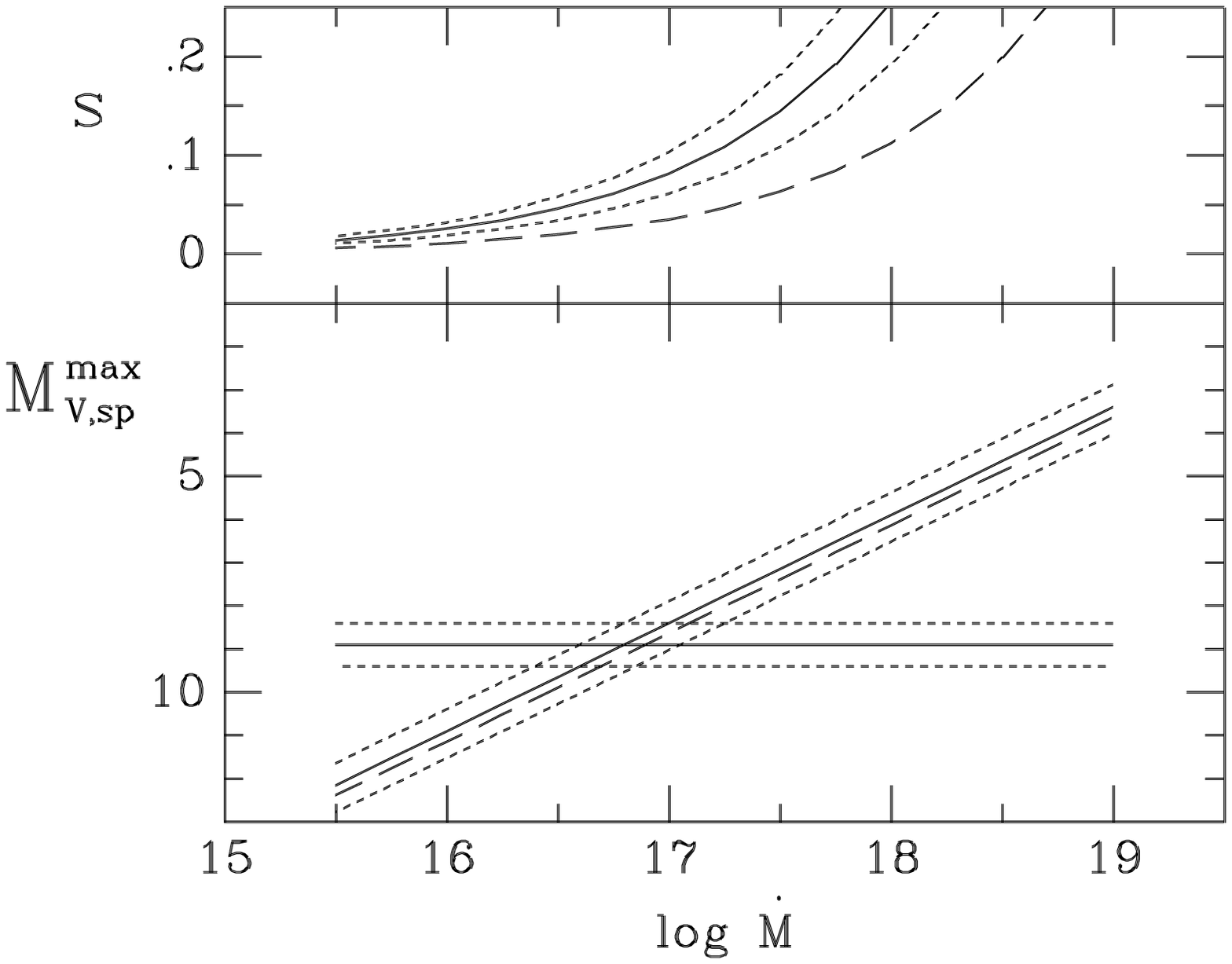} 
\vskip 5truemm
\FigCap { The $M_{V,sp}^{max}-\dot M_{tr}$ ({\it bottom}) and $s-\dot M_{tr}$ 
({\it top}) relations calculated from Eqs.(5)-(9) with $\eta=0.5$.  
The solid and dotted lines correspond to $r_d=(0.7\pm0.1)~r_{Roche}$ 
and $\log T_{sp}=3.82$. The broken lines -- to $r_d=0.7~r_{Roche}$ and 
$\log T_{sp}=4.00$. The horizontal lines in the lower plot represent the absolute 
visual magnitude of the spot at maximum (Section 2.2): $M_{V,sp}^{max}=8.9\pm0.5$. 
}
\end{figure}

\subsection { Comments on $\dot M_{tr}=(1.1-3.8)\times 10^{18}$ g/s}

Such a value was obtained by Schreiber and Lasota (2007) from their estimates 
involving the amount of mass accreted during outburst. 
Apart from our direct determination of the much lower value of $\dot M_{tr}$ 
(Section 3.3) there are several other arguments which imply that such a high 
mass transfer rate is simply impossible. 

From Fig.1 we find that the luminosity of the hot spot (at maximum) 
corresponding to $\dot M_{tr}=(1.1-3.8)\times 10^{18}$ g/s should be 
$M_{V,sp}^{max}=5.3\pm 1.3$. If so, the contribution from the spot would 
not only dominate in the shape of the light curve but would also increase 
the total luminosity of the system well above the observed $M_V=5.6\pm0.5$. 

From Fig.1 we also find that at $\dot M_{tr}=(1.1-3.8)\times 10^{18}$ g/s 
the dimensions of the spot should be larger, or even much larger than 
$s\sim0.2$, which is simply unrealistic. 

In addition, with $\dot M\sim 2\times 10^{18}$ g/s, the amount of mass added 
to the disk by the quiescent stream would be $\Delta M_D\sim 9\times 10^{24}$ g. 
However, the total mass of the disk obtained from dwarf nova model calculations, 
intended to represent SS Cyg and Z Cam (Hameury et al. (1998, Fig.10; 
Buat-M{\'e}nard 2001, Fig.2) is only $M_D\sim (1.0-2.5)\times 10^{24}$ g. 
The value of $\Delta M_D\sim 9\times 10^{24}$ g would then be significantly 
larger than the total mass of the disk. 
(At this point it may also be worth to mention that $M_{D,max}$ discussed 
by Schreiber and Lasota [2007, Eq.5], and used by them to estimate $\dot M_{tr}$, 
is {\it not} the mass of the disk at outburst maximum, but an {\it absolute 
upper limit} to disk mass just before the outburst).

\subsection { The Mass Transfer Rate at Quiescence }

Using the $M_{V,sp}^{max}-\dot M_{tr}$ relation presented in Fig.1 
and $M_{V,sp}^{max}=8.9\pm0.5$ (Section 2.2) we obtain  

\beq
\log \dot M_{tr}\approx 16.8\pm 0.3~.  
\eeq

\noindent 
This is the main result of our analysis. It shows that the mass transfer rate 
in SS Cyg at quiescence is safely below the critical mass transfer rate 
(Eqs.3 in Section 2.3). 
In particular, this is true even in the case of the much lower "revised" 
value of $\log~\dot M_{crit}^{3.65}=17.2$. 

The radius of the spot at $\log \dot M_{tr}\approx 16.8\pm 0.3$ 
(see Fig.1--{\it top}) is $s\sim 0.05$ which looks reasonable: 
it is only slightly larger than in other well studied dwarf novae with 
comparable orbital periods, e.g. U Gem (Smak 1996), or IP Peg (Smak 2001). 

From the mass transfer rate and the duration of the cycle we can also calculate 
the amount of mass added to the disk by the quiescent stream:  
$\Delta M_D\sim 3\times 10^{23}$ g. 
Comparing this with $M_D\sim (1.0-2.5)\times 10^{24}$ g (see Section 3.2) 
we conclude that it represents reasonable fraction -- roughly 10-30 percent 
-- of the total mass of the disk. 

At this point we recall that our estimate of the luminosity of the spot 
(Section 2.2), based on the light curves published by Voloshina and Khruzina 
(2000ab), was rather crude. It is now re-assuring to note that 
the selfconsistency of results presented above provides an independent 
argument in favor of the adopted value of $M_{V,sp}^{max}=8.9\pm0.5$.

\subsection { The Mass Transfer Rate during Outbursts }

The accretion rate during outburst maxima was estimated by Schreiber and Lasota 
(2007, Eq.3) as $\dot M_{out}\sim 9\times 10^{18}$ g/s. 
Combined with our value of $\dot M_{tr}\sim 6\times 10^{16}$ g/s it implies 
that the mass transfer rate during outbursts is enhanced by -- very roughly -- 
factor of $\sim$100. Considering all uncertainties involved we note that this is 
similar to earlier estimates for U Gem (Smak 2005) and for SU UMa type dwarf 
novae (Smak 2004).

\section { Conclusion } 

Results presented above imply that the answer to the question posed 
by Schreiber and Lasota (2007) in the title of their paper is quite simple: 
Nothing is wrong with SS Cyg, nor with the theory of dwarf nova outbursts.

\begin {references} 


\refitem {Bitner, M.A., Robinson, E.L., Behr, B.B.} {2007} {\ApJ} {662} {564}

\refitem {Bruch, A., Engel, A.} {1994} {\AA Suppl.} {104} {79} 

\refitem {Buat-M{\'e}nard, V., Hameury, J.-M., Lasota, J.-P.} {2001} 
          {\AA} {369} {925}

\refitem {Hameury, J.-M., Menou, K., Dubus, G., Lasota, J.-P., Hur{\'e}, J.-M.}
          {1998} {\MNRAS} {298} {1048}

\refitem {Harrison, T.E., McNamara, B.J., Szkody, P., McArthur, B.E., 
          Benedict, G.F., Klemola, A.R., Gilliland, R.L.} {1999} 
          {\ApJ Letters} {515} {L93}

\refitem {Kurucz, R.L.} {1993} {\rm CD-ROM, No.13} {~} {~}

\refitem {Lasota, J.-P.} {2001} {New Astronomy Reviews} {252} {100}

\refitem {Paczy{\'n}ski, B., Schwarzenberg-Czerny, A.} {1980} {\Acta} {30} {127} 

\refitem {Schreiber, M.R., G{\"a}nsicke, B.T.} {2002} {\AA} {382} {124}

\refitem {Schreiber, M.R., Lasota, J.-P.} {2007} {\AA} {473} {897}

\refitem {Smak,J.} {1996} {\Acta} {46} {377}   

\refitem {Smak,J.} {2001} {\Acta} {51} {279}   

\refitem {Smak,J.} {2002} {\Acta} {52} {429}   

\refitem {Smak,J.} {2004} {\Acta} {54} {221}   

\refitem {Smak,J.} {2005} {\Acta} {55} {315}   

\refitem {Voloshina, I.B., Khruzina, T.S.} {2000a} {\it Astron.Zh.} {77} {109}

\refitem {Voloshina, I.B., Khruzina, T.S.} {2000b} {\it Astron.Rep.} {44} {89}

\end {references}

\end{document}